\documentclass[10pt,preprint]{aastex62}

\graphicspath{{plots/}}

\usepackage{mathrsfs, amsmath}
\usepackage{dblfloatfix}
\usepackage{csquotes}
\usepackage{dblfloatfix}
\usepackage{fixltx2e}
\usepackage{url}
\usepackage{natbib}

\usepackage{graphicx}
\usepackage{enumitem}

\newcommand{\nref}{$N_\text{ref}$}
 
\received{February 18, 2019}
\revised{April 26, 2019}
\revised{May 9, 2019}
\accepted{May 14, 2019}
\submitjournal{AJ}

\shorttitle{SCExAO/CHARIS chromaticity analysis}
\shortauthors{Gerard, B. L. et al.}

\begin{document}
\title{A Chromaticity Analysis and PSF Subtraction Techniques for SCExAO/CHARIS data}

\correspondingauthor{Benjamin L. Gerard}
\email{bgerard@uvic.ca}

\author[0000-0003-3978-9195]{Benjamin L. Gerard}
\affil{University of Victoria, Department of Physics and Astronomy, 3800 Finnerty Rd, Victoria, V8P 5C2, Canada}
\affiliation{National Research Council of Canada, Astronomy \& Astrophysics Program, 5071 West Saanich Rd, Victoria, V9E 2E7, Canada}

\author[0000-0002-4164-4182]{Christian Marois}
\affiliation{National Research Council of Canada, Astronomy \& Astrophysics Program, 5071 West Saanich Rd, Victoria, V9E 2E7, Canada}
\affiliation{University of Victoria, Department of Physics and Astronomy, 3800 Finnerty Rd, Victoria, V8P 5C2, Canada}

\author[0000-0002-7405-3119]{Thayne Currie}
\affiliation{NASA-Ames Research Center, Moffett Field, CA, USA}
\affiliation{Subaru Telescope, National Astronomical Observatory of Japan, 650 North A'oh\={o}k\={u} Place, Hilo, HI 96720, USA}
\affiliation{Eureka Scientific, 2452 Delmer Street Suite 100, Oakland, CA, USA}

\author{Timothy Brandt}
\affiliation{Department of Physics, University of California-Santa Barbara, Santa Barbara, CA, USA}

\author[0000-0001-6305-7272]{Jeffrey K. Chilcote}
\affiliation{Department of Physics, University of Notre Dame, 225 Nieuwland Science Hall, Notre Dame, IN, 46556, USA}

\author{Zachary H. Draper}
\affil{University of Victoria, Department of Physics and Astronomy, 3800 Finnerty Rd, Victoria, V8P 5C2, Canada}
\affiliation{National Research Council of Canada, Astronomy \& Astrophysics Program, 5071 West Saanich Rd, Victoria, V9E 2E7, Canada}

\author[0000-0001-5978-3247]{Tyler Groff}
\affiliation{NASA-Goddard Space Flight Center, Greenbelt, MD, USA}

\author{Olivier Guyon}
\affiliation{Steward Observatory, University of Arizona, Tucson, AZ 85721, USA}
\affiliation{Astrobiology Center, National Institutes of Natural Sciences, 2-21-1 Osawa, Mitaka, Tokyo, Japan}

\author{Masahiko Hayashi}
\affiliation{NAOJ, 2-21-1 Osawa, Mitaka, Tokyo 181-8588, Japan}

\author{Nemanja Jovanovic}
\affiliation{Department of Astronomy, California Institute of Technology, 1200 E. California Blvd., Pasadena, CA 91125, USA}

\author{Gillian R. Knapp}
\affiliation{Department of Astrophysical Science, Princeton University, Peyton Hall, Ivy Lane, Princeton, NJ 08544, USA}

\author{Tomoyuki Kudo}
\affiliation{Subaru Telescope, National Astronomical Observatory of Japan, 650 North A'oh\={o}k\={u} Place, Hilo, HI 96720, USA}

\author{Jungmi Kwon}
\affiliation{ISAS/JAXA, 3-1-1 Yoshinodai, Chuo-ku, Sagamihara, Kanagawa 252-5210, Japan}

\author[0000-0002-3047-1845]{Julien Lozi}
\affiliation{Subaru Telescope, National Astronomical Observatory of Japan, 650 North A'oh\={o}k\={u} Place, Hilo, HI 96720, USA}

\author{Frantz Martinache}
\affiliation{Universit\'{e} C\^{o}te d'Azur, Observatoire de la C\^{o}te d'Azur, CNRS, Laboratoire Lagrange, France}

\author[0000-0003-0241-8956]{Michael McElwain}
\affil{NASA-Goddard Space Flight Center, Greenbelt, MD, USA}

\author{Motohide Tamura}
\affil{Department of Astronomy, Graduate School of Science, The University of Tokyo, 7-3-1, Hongo, Bunkyo-ku, Tokyo, 113-0033, Japan}
\affil{Astrobiology Center, National Institutes of Natural Sciences, 2-21-1 Osawa, Mitaka, Tokyo, Japan}

\author{Taichi Uyama}
\affil{Department of Astronomy, Graduate School of Science, The University of Tokyo, 7-3-1, Hongo, Bunkyo-ku, Tokyo, 113-0033, Japan}

\begin{abstract}
We present an analysis of instrument performance using new observations taken with the Coronagraphic High Angular Resolution Imaging Spectrograph (CHARIS) instrument and the Subaru Coronagraphic Extreme Adaptive Optics (SCExAO) system. In a correlation analysis of our datasets (which use the broadband mode covering J through K band in a single spectrum), we find that chromaticity in the SCExAO/CHARIS system is generally worse than temporal stability. We also develop a point spread function (PSF) subtraction pipeline optimized for the CHARIS broadband mode, including a forward modelling-based exoplanet algorithmic throughput correction scheme. We then present contrast curves using this newly developed pipeline. An analogous subtraction of the same datasets using only the H band slices yields the same final contrasts as the full JHK sequences; this result is consistent with our chromaticity analysis, illustrating that PSF subtraction using spectral differential imaging (SDI) in this broadband mode is generally not more effective than SDI in the individual J, H, or K bands. In the future, the data processing framework and analysis developed in this paper will be important to consider for additional SCExAO/CHARIS broadband observations and other ExAO instruments which plan to implement a similar integral field spectrograph broadband mode.
\end{abstract}
\section{Introduction}
\label{sec: intro}
Ground-based exoplanet imaging is currently uniquely poised to understand how giant planets form and evolve over time. Distinct from other exoplanet detection methods, exoplanet imaging is currently sensitive to wide separations (greater than about 10 AU), young ages (less than about 1 Gyr), and high masses (greater than about one Jupiter mass; e.g., \citealt{gpies}). However, the push to closer separations, older ages, and lower masses is constantly improving with new instrumentation and processing algorithms.

Post-processing of high contrast imaging data is currently a crucial component of reaching the necessary contrasts in order to detect new exoplanets by subtracting the correlated speckle noise around their host star. There are three main point spread function (PSF) subtraction observing strategies used to improve the exoplanet signal to noise ratio (S/N): spectral differential imaging \cite[SDI;][]{ssdi0,ssdi1,ssdi2,ssdi3}, angular differential imaging \citep[ADI;][]{adi}, and the commonly used reference differential imaging RDI algorithm. Although these strategies are often combined with more advanced least-squares subtraction algorithms \citep[e.g.,][]{loci, klip}, all three methods remain fundamentally limited by PSF chromaticity (SDI and RDI) and/or temporal stability (ADI and RDI). Thus, understanding and mitigating the level of chromatic and temporal decorrelation in high contrast imaging datasets is a crucial step towards improving contrast to detect fainter and/or lower mass exoplanets. 

Although significant efforts have been made to understand the relative impact of temporal stability vs. chromaticity from a numerical perspective \citep[e.g.][]{gpi_fresnel, sphere_stability}, a similar comprehensive on-sky analysis with second generation high contrast imaging instruments that are equipped with an integral field spectrograph (IFS) has not yet been completed. \cite{cor1}, and \cite{cor2} have carried out temporal analyses on the quasi-static speckle lifetime, but with no corresponding chromaticity analysis. Following from the work of \cite{sosie}, temporal correlation analyses have informed some new ADI+SDI PSF subtraction pipelines to select reference images ranked by correlation \citep[e.g.][]{aloci,pyklip}. \cite[][]{tloci} presented an analyses of time- vs. wavelength-dependent correlation in individual target images. In this paper we will extend this analysis to multiple target images; later we will show that such correlation can vary significantly depending on observing conditions and on location in the coronagraphic image.

The main focus of this paper is thus to present a detailed analysis of chromaticity and temporal stability using our observations from the Subaru adaptive optics system (AO188; \citealt{ao188}),  Subaru Coronagraphic Extreme Adaptive Optics (SCExAO; \citealt{scexao}), and Coronagraphic High Angular Resolution Imaging Spectrograph (CHARIS; \citealt{charis1,charis2}). We also present new modifications to the ADI+SDI-based template locally optimized combination of images \citep[TLOCI;][]{tloci} PSF subtraction pipeline for these datasets. Our SCExAO/CHARIS datasets use a Lyot coronagraph in the broadband JHK mode, covering the full 1.15 - 2.39 $\mu$m range at R$\sim$18 \citep{charis1}; this new low resolution broadband mode, currently unique to CHARIS amongst current high contrast imaging instruments, is thought to provide a potential gain in achievable contrast through SDI PSF subtraction for an achromatic system \citep{gpi_fresnel, tloci}, and so our performance analysis and PSF subtractions methods will focus on impacts of using this broadband mode. 

The structure of this paper is as follows: in \S\ref{sec: obs} we present details of our target observations, in \S\ref{sec: data_analysis} we present the architecture of our PSF subtraction pipeline (\S\ref{sec: psf_subt}) and then use the same reference image selection and correlation calculation procedures to carry out a detailed chromaticity analysis (\S\ref{sec: corr}), in \S\ref{sec: results} we present final contrast curves and discuss instrument performance implications for the future, and we conclude in \S\ref{sec: conclusions}. Additional correlation analysis is presented in appendix \ref{sec: cor_cut}.
\section{Observations}
\label{sec: obs}
We observed a handful of young, nearby two temperature-component debris disks, the extrasolar analogues of the Asteroid and Kuiper belts in our Solar System \citep{wyatt_rev}. We identified targets in \cite{two_comp} that had not yet been observed by the Gemini Planet Imager (GPI; \citealt{gpi}) or Spectro-Polarimetric High-contrast Exoplanet REsearch instrument (SPHERE; \citealt{sphere}) and/or are inaccessible to either for being too far north. We also observed two additional targets in 2018A through ancillary time \citep{klauss_time}. Our observed targets are listed in Table \ref{tab: targets}, including dates of observations, exposure times, and field of view (FOV) rotation.
\begin{table}[!hb]
\begin{tabular*}{\textwidth}{c @{\extracolsep{\fill}} ccccc}
UT date & name & $m_V$ & $t_\text{exp}$ & $n_\text{frames}$ & $\Delta_\text{PA}$ \\
(d/m/yy) & (HD) & & (s) & & (deg) \\
\hline
9/4/17 & 70313 & 5.51 & 15 & 107 & 25 \\
9/4/17 & 98673 & 6.43 & 31 & 91 & 22 \\
9/4/17 & 109085 & 4.3 & 13 & 93 & 18 \\ 
9/4/17 & 125162 & 4.18 & 20 & 93 & 21 \\
9/4/17 & 141378 & 5.53 & 31 & 70 & 21 \\
7/1/18 & 23267 & 6.88 & 60 & 50 & 32 \\
8/1/18 & `` '' & `` '' & `` '' & 40 & 24 \\
8/1/18 & 38206 & 5.73 & 31 & 118 & 27 \\
\end{tabular*}
\caption{Table of our target observing parameters.}
\label{tab: targets}
\end{table}

Observations were made in the standard pupil tracking mode for ADI. The coronagraphic setup in this broadband mode utilized the standard Lyot coronagraph, applying the 217 mas diameter focal plane mask (FPM), for which the inner working angle (IWA; separation at which an off-axis companion would have 50\% optical throughput relative to the center of the star) is 113 mas; at the time, no other coronagraph options offered sufficient performance over the full broadband mode of CHARIS. The deformable mirror waffle spots \citep{marois06, sivaramakrishnan, nem_sat_mod}, or ``satellite spots,'' are applied with a non-standard amplitude for the April 2017 data, and so instead an on-sky non-coronagraphic data sequence (using an ND filter but with the same Lyot stop) is taken to measure the star-to-spot ratio directly (see \S\ref{sec: drp}). The January 2018 data uses the standard 50 nm amplitude spots and corresponding pre-computed star-to-spot ratio as in \cite{scexao_spot_ratios}.
\section{Data Analysis}
\label{sec: data_analysis}
\subsection{Data Cube Extraction and Setup}
\label{sec: drp}
We used the CHARIS data reduction pipeline \citep[DRP;][]{charis_drp} to generate data cubes from recorded IFS detector images, which uses a least-squares method to extract the flux in each slice of the data cube \citep{zack_least_squares}. A background subtraction, flat field correction, and bad pixel mask are also applied in the extraction process as described in \cite{charis_drp}. All images are then prepared for subsequent PSF subtraction (\S\ref{sec: psf_subt}) and correlation analysis (\S\ref{sec: corr}). The satellite spots are used for image registration, magnification with wavelength to align speckles, flux-normalization to remove the wavelength dependence of the broadband transmission filter and stellar spectrum, and contrast calibration.  We chose to magnify the images with wavelength centered around slice 16 of 22 ($\lambda_0=2.0 \mu$m). Slice 16 was chosen instead of slice 11 ($\lambda=1.7 \mu$m; in the middle of the bandpass) because of wavelength-dependent sampling issues. The CHARIS PSF in J and H band is slightly under-sampled, and we found that cubic spline interpolation, used to rotate and magnify the images for our ADI+SDI PSF subtraction algorithm (\S\ref{sec: psf_subt}), provided a more accurate centroiding precision from numerical interpolation in K band. The four satellite spots all lie at 15.91 $\lambda_0/D$ separation, and so measured satellite spot positions in the registered, magnified images are used to calculate a plate scale for contrast curve measurements (\S\ref{sec: results}). We were not able to use any on-sky astrometric data to validate these derived plate scale values. Images are then high-pass-filtered using a $4.1\times4.1\; \lambda_0/D$ median boxcar filter to attenuate the smooth AO halo while retaining a close-to-unity algorithmic throughput of $\lambda/D$-scale static speckles and any possible planet(s). The $\lambda^2$ dependence of the satellite spot intensity is also removed during the flux normalization procedure. The satellite spots are omitted from reference image selection in \S\ref{sec: psf_subt} and correlation calculations for analysis in \S\ref{sec:  corr}. Finally, all registered, magnified, flux-normalized, high-pass-filtered images are converted into ``raw'' contrast units (i.e., the pixel value-to-star ratio throughout the whole image) using measured and pre-calibrated star-to-satellite spot ratios, whose data acquisition is described at the end of \S\ref{sec: obs}. The same registration, magnification, flux normalization, and high-pass filtering procedure is applied to the custom April 2017 star-to-spot calibration data before computing the ratios.
\subsection{PSF Subtraction}
\label{sec: psf_subt}
\begin{figure}[!h]
\centering
\includegraphics[width=0.7\textwidth]{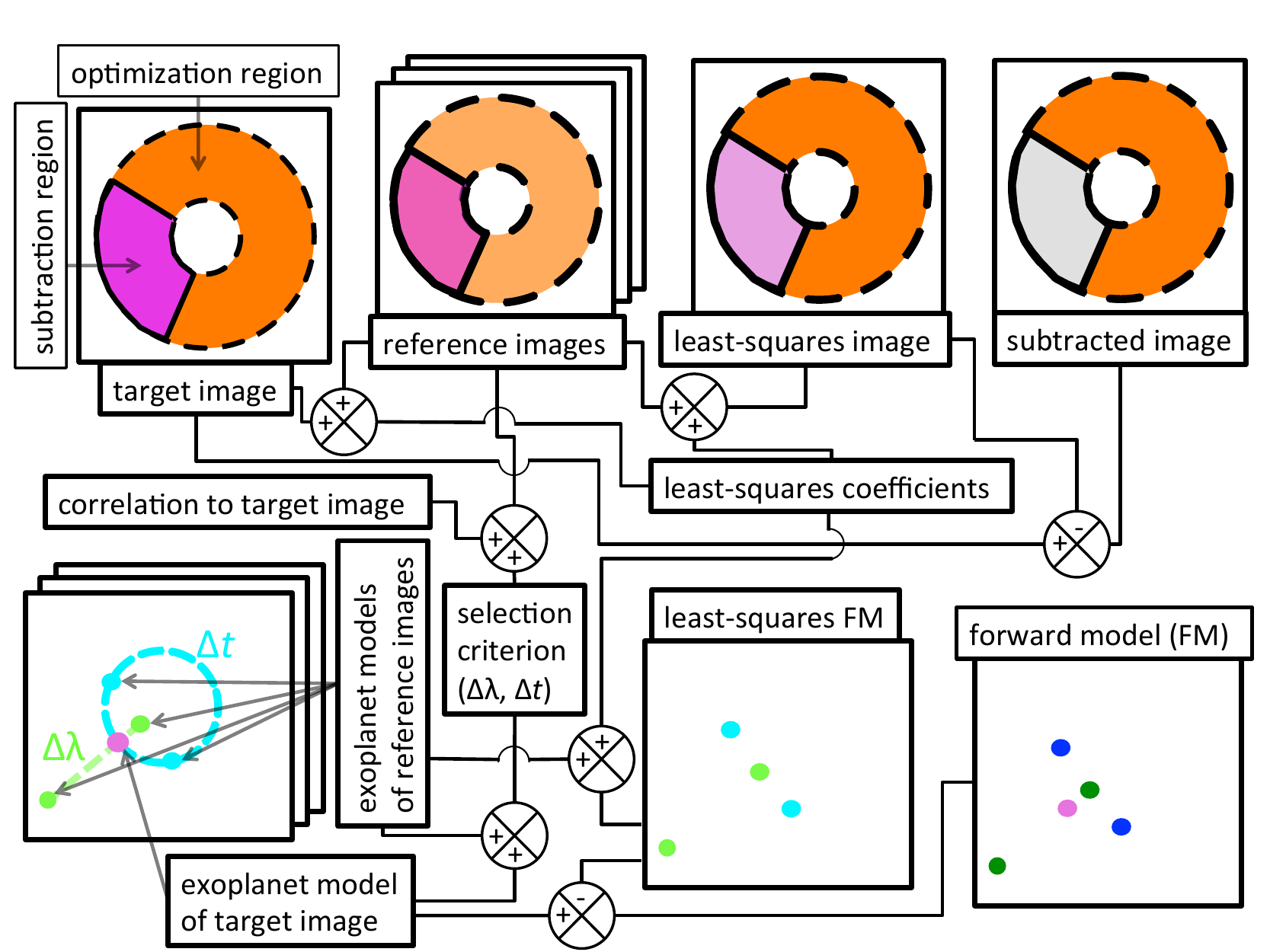}
\caption{A block diagram illustrating the process of least-squares-based PSF subtraction and forward modelling. The junctions combine the two boxes (connected to a ``+'' or ``-'') and output to a third box. The junctions with a ``-'' sign indicate that the output box is the difference between the two images represented by input boxes. All other boxes, where the two junctions are ``+'' signs, represent a more abstract process of combining two input concepts  (i.e., an image, a vector of images, a vector of coefficients, and/or a selection criterion) to produce an output concept. The inputs, on the lower left of the diagram, include a model of both the exoplanet target image and all other reference images in the sequence, both in time and wavelength, including the assumed spectrum of the planet. These models are then used to define a selection criterion, which is combined with correlation of each reference image in the sequence to the target image in order to select the best reference images to use in subtracting the target image. After a few intermediate steps involving the calculation of least squares coefficients, the outputs are illustrated on the right side of the diagram, including a subtracted image along with a noiseless forward model of the planet signal in the subtracted image.}
\label{fig: tloci}
\end{figure}

Our ADI+SDI PSF subtraction algorithms build on the least-squares-based locally optimized combination of images \citep[LOCI;][]{loci} algorithm, further developed by \cite{sosie} and \cite{tloci} as Template LOCI (TLOCI). A conceptual illustration of the procedure used in our TLOCI pipeline to subtract a single region of a single target image is shown in Fig. \ref{fig: tloci}.
\begin{figure}[!h]
\centering
\includegraphics[width=0.6\textwidth]{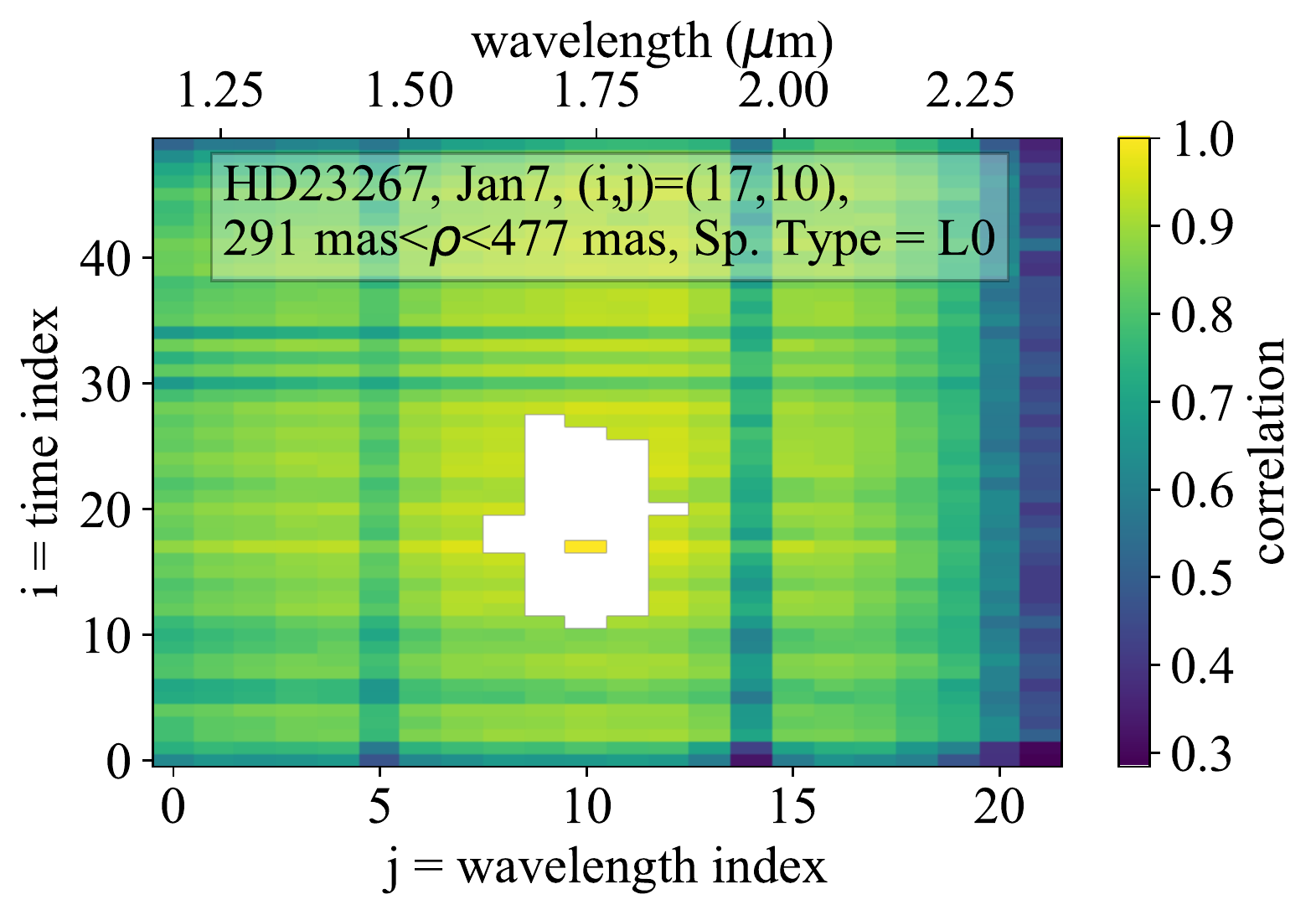}
\caption{Correlation of selectable reference images to a single target image. The target image is illustrated with a correlation of 1 in the center of the white region. Various correlation trends illustrate the stability of the sequence in time and wavelength. Darker vertical lines illustrate wavelength-dependent effects, including the atmospheric water bands at wavelength slices 5 and 14, a degradation of performance towards the red edge of K band, while darker horizontal lines illustrate time-dependent effects from variable AO  and/or quasi-static stability.}
\label{fig: cor1}
\end{figure}

In addition to the process of generating least-squares coefficients to subtract a region of the target image, Fig. \ref{fig: tloci} also illustrates the principle of forward modelling (FM), where the same coefficients are also used to model the algorithmic exoplanet throughput during the subtraction. As shown, a selection criterion, or \textit{aggressiveness}, specifies that reference images be selected beyond a radial and/or azimuthal gap in wavelength and/or time, respectively, from the target image in order to minimize algorithmic exoplanet throughput losses. The highest aggressiveness of 1 (using dimensionless units calculated from relative aperture photometry between the target and reference image models) would allow using any reference image in the full target sequence, including the target image. In contrast, the lowest aggressiveness of 0 would not allow any reference images in the target sequence to be selected. Aggressiveness is also illustrated in Fig. \ref{fig: cor1}. In this figure, a range of times and wavelengths around the target image are not selectable because, if used, a planet present in the target image would be partially subtracted by the same planet in the reference images. This un-selectable region of Fig. \ref{fig: cor1}, illustrated by a white color, would be either larger or smaller for a lower or higher value of user-defined aggressiveness, respectively. The size of this un-selectable region  also depends on the separation in the image, assumed exoplanet spectrum, wavelength, and parallactic angle relative to the target image. The target image is highlighted in the center of this region with a correlation of one. Then, reference images are chosen from the remaining sample based on correlation, also illustrated in Fig. \ref{fig: cor1}. Generally, the most correlated reference images are near the target image time or wavelength, although this depends on the temporal stability and chromaticity of the sequence. A more correlated set of reference images (i.e., those with a more similar speckle pattern to the target image) provides a better basis for the least-squares algorithm and will yield a deeper final contrast without changing any other free parameters \citep{tloci}. We will examine the chromatic and temporal dependence of these correlations further in \S\ref{sec: corr}. Additional details are described below about the setup and new modifications of our TLOCI pipeline.

To help define, identify, and remove uncorrelated images from each dataset, we will introduce a new free parameter into our PSF subtraction pipeline called a correlation cut. For a single target image, the value of a correlation cut will define the minimum level of correlation that is rendered acceptable in a reference image used to subtract the target image speckle pattern. A given target image/region will be discarded from the subtracted sequence if there are not at least \nref\; reference images above the correlation cut value, all of which must pass a selection criterion, where \nref\; and aggressiveness are also free parameters. For example, if correlation cut = 0.5 and \nref\;= 10 but there are only five reference images that are greater than 50\% correlated to the target image using a moderate aggressiveness, the target image/region will be discarded from the sequence. The target image would be discarded in this case even if, e.g., one of those five reference images is 99\% correlated to the target image. Thus, it is clearly important to use a physically motivated correlation cut so that a maximal number of target images in an observing sequence can be subtracted while still rejecting a maximal number of uncorrelated reference images. \cite{sosie} and subsequent papers have shown that using correlated reference images is crucial to prevent unnecessary noise propagation in the covariance matrix inversion step of a least-squares/principal component analysis subtraction algorithm. The correlation cut values are dimensionless by definition and have a minimum and maximum of 0 and 1, respectively. Analysis in appendix \ref{sec: cor_cut} inform the decision to use the following correlation cut values for the four different radial sections of an image, progressing from innermost to outermost separation: 0.8, 0.74, 0.67, and 0.59. Although the specific angular separations of these regions varies for each target based on our method of measuring the pixel scale using the satellite spots (\S\ref{sec: drp}), the average separation and standard deviation across all targets is: 104$\pm$2 mas $<\rho<$ 286$\pm$4 mas, 286$\pm$4 mas $<\rho<$ 468$\pm$7 mas, 468$\pm$7 mas $<\rho<$ 649$\pm$10 mas, and $\rho>$ 649$\pm$10 mas. A graphical illustration of these zones is illustrated later in \S\ref{sec: psf_subt}. Note that we chose to fix the radial separation of subtraction regions based on pixel separation, which is why the derived angular separation varies slightly from target to target. In pixels, the radial separations for each annulus for all targets are  $7<\rho<19$, $19<\rho<32$, $32<\rho<44$, and $\rho>44$.

Based on the definitions and work in \cite{sosie} and \cite{mara_thesis}, for all subsequent analyses in this paper we use an L0 or T6 input spectrum (J02281101+2537380 and J02281101+2537380 from \citealt{l0} and \citealt{t6}, respectively, subsequently convolved and binned to the appropriate CHARIS spectral resolution), an aggressiveness of 0.5 and 15 reference images per least-squares subtraction. \cite{me_gpi} found, on GPI datasets, that optimizing aggressiveness and number of reference images yielded a negligible gain in contrast, and so here we use the same aggressiveness, number of reference images per subtraction, and input spectral types (to define a selection criterion and for subsequent forward modelling). We use a singular value decomposition (SVD) cut-off value for the covariance matrix inversion of $1\times10^{-4}$, as in \cite{me_gpi}, where a maximum value of 1 would set all the LOCI coefficients to zero. Although this is a low value compared to other algorithms \citep[e.g.][]{kappa_and}, note that there are many degeneracies between the number of reference images used, reference image correlation, and SVD cut-off \citep{sosie}; our approach here uses a lower SVD cut-off but compensates by selecting more correlated reference images.

\begin{figure}[!h]
\centering
\includegraphics[width=1.0\textwidth]{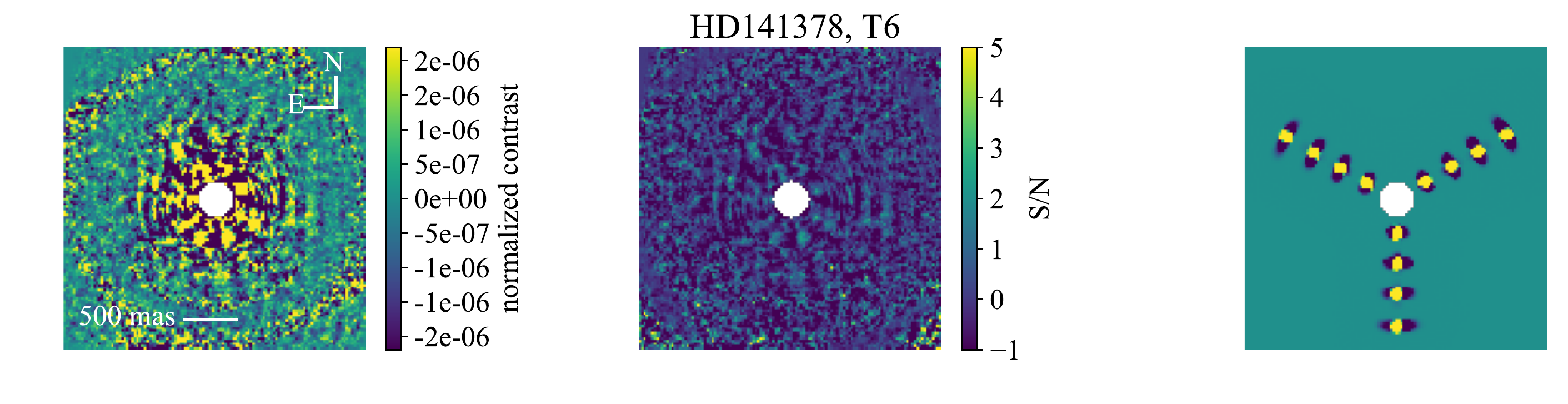}
\caption{PSF-subtracted data products for a single target, HD141378. Left: an image of a collapsed PSF-subtracted sequence. Middle: an S/N map of the left image. Right: a collapsed forward model map. Note the relative absence of dark shadows in the radial relative to azimuthal direction. This suggests that images at the same wavelength and different time are more correlated than images at a different time and same wavelength, causing the former to be selected as reference images more often than the latter.}
\label{fig: psf_subt}
\end{figure}

In a new addition to our TLOCI pipeline, the central region of each subtraction region is defined by both the parallactic angle and wavelength at any given timestamp, and so the location of a single region will both rotate with time and magnify with wavelength over a single observing sequence, similar to the approach in \citet{pyklip}. These definitions are used so that the subtraction regions line up after de-rotation and de-magnification, which allows the generation of a FM map for each target. The LOCI optimization regions are defined for each subtraction region as the other two azimuthal regions at the same radial separation. This subtraction/optimization region geometry ensures that the least-squares coefficients are never generated from the subtraction zone where a planet would be. S/N maps are calculated using back-rotated images, as in \cite{me_gpi}. A few example data products of a single target sequence from our PSF-subtraction pipeline are shown in Figure \ref{fig: psf_subt}, including a subtracted collapsed data cube, corresponding S/N map, and corresponding FM map.

\begin{figure}[!h]
\begin{center}
\includegraphics[width=1.0\textwidth]{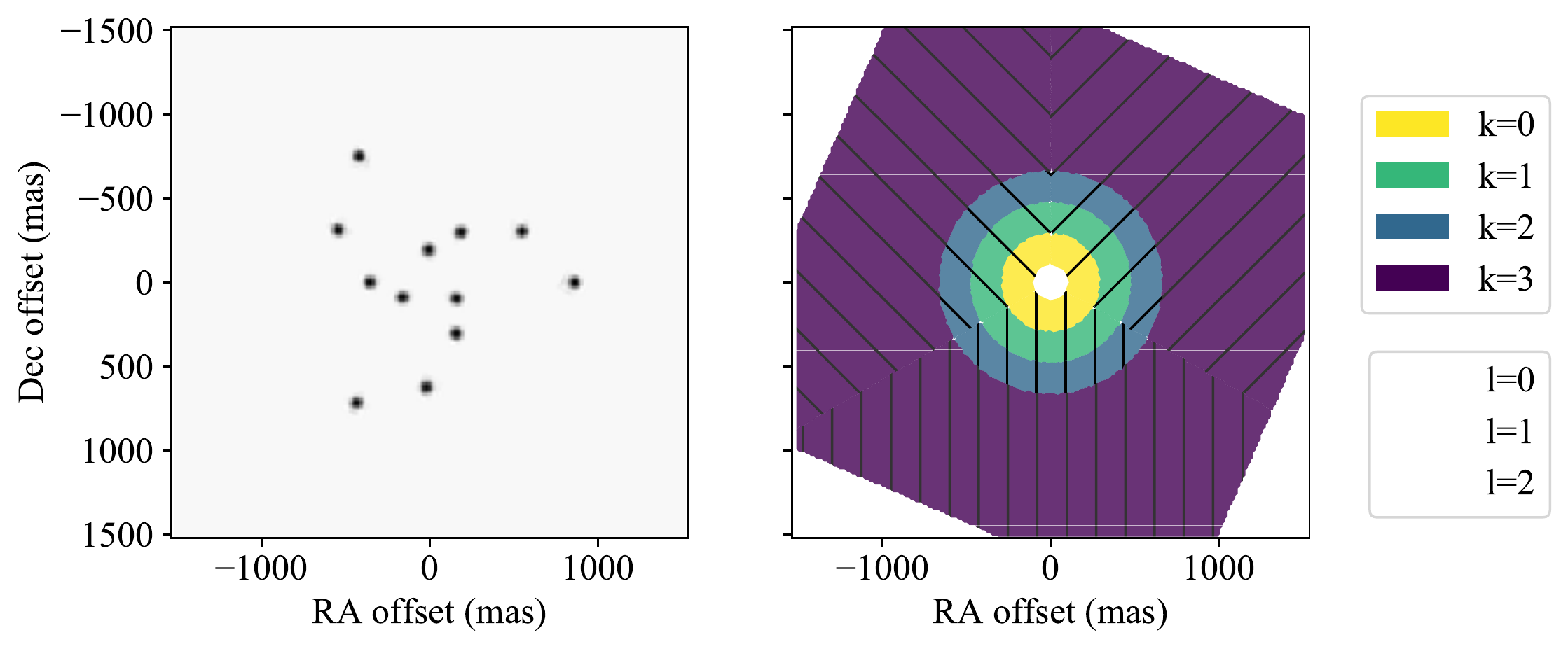}
\end{center}
\caption{Right: definitions of the different subtraction regions defined in the final collapsed image. ``k'' and ``l'' represent the radial and azimuthal indices for each subtraction region, respectively. Left: the locations of simulated planets (one per subtraction).}
\label{fig: boot_reg}
\end{figure}

\begin{figure}[!h]
\begin{center}
\includegraphics[width=1.0\textwidth]{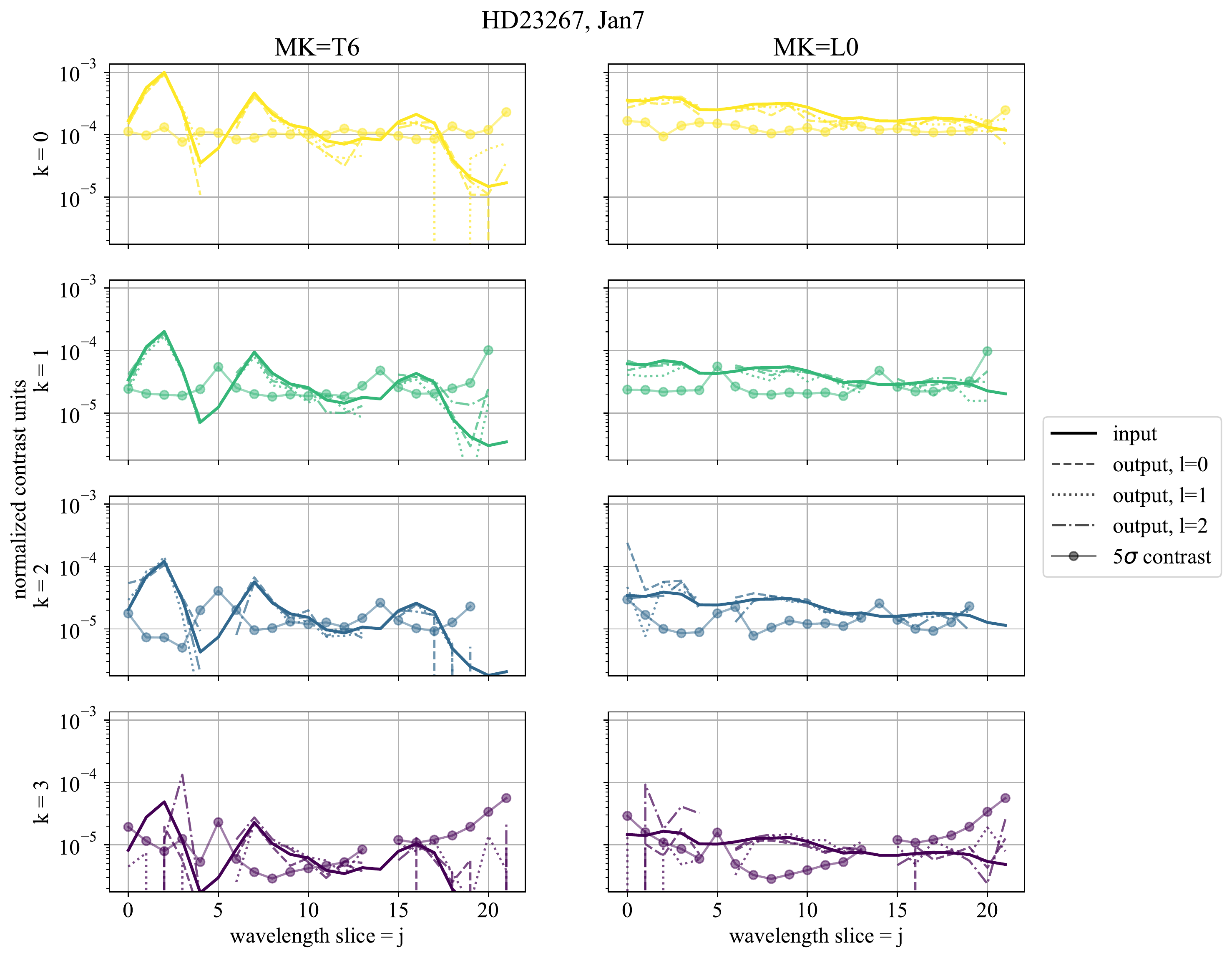}
\end{center}
\caption{The injected and recovered fluxes of the simulated planets illustrated in Fig. \ref{fig: boot_reg}, as well as the 5$\sigma$ contrasts corresponding to the separation of each simulated planet (calculated from the back-rotated images). The same conventions for ``k'' and ``l'' are used here as in Fig. \ref{fig: boot_reg}.}
\label{fig: boot}
\end{figure}

For each target we also run a bootstrapping procedure, illustrated in Figures \ref{fig: boot_reg} and \ref{fig: boot}. 
As in \cite{sosie} and \cite{tloci}, we implement a FM algorithmic throughput correction procedure in each subtraction region of the image, including a model for the radial variation of the algorithmic throughput within each region. In addition to the center, four additional FM throughput corrections using the same LOCI coefficients are computed at smaller and larger separations within each subtraction region. A fifth order polynomial is then fit to the radial variation of throughput correction values to generate a continuous function across the region. As a result of this throughput correction procedure, a bootstrapped planet at any location in the image (i.e., not necessarily at the radial or azimuthal center of a subtraction region), is extracted at the input value and does not need any additional throughput correction. This is illustrated in Fig. \ref{fig: boot}. The simulated planets were injected throughout the image, one per subtraction region, such that none lies at the radial or azimuthal center of the region, as in the left panel of Fig. \ref{fig: boot_reg}. Almost all recovered fluxes match the input value within their expected scatter, materialized by the 5$\sigma$ contrast value at the location of the planet.

\subsection{Chromaticity Analysis}
\label{sec: corr}
As discussed in \S\ref{sec: intro}, in a perfectly achromatic system, the larger ``lever arm'' of wavelength coverage over the JHK filter of CHARIS should provide a better algorithmic exoplanet throughput than a narrower bandpass, and so if both cases provide the same level of speckle subtraction, the final throughput-corrected contrast should be better in the former case. However, in reality broadband systems are not achromatic, and the level of wavefront decorrelation, or chromaticity (i.e., separate from wavelength magnification), will ultimately determine the limits of speckle subtraction via SDI processing \citep[e.g.,][]{gpi_fresnel}. In this section we will measure the chromaticity in our CHARIS datasets, which will then inform the SDI speckle subtraction limits in \S\ref{sec: results}.

\begin{figure}[!ht]
\begin{center}
\includegraphics[width=1.0\textwidth]{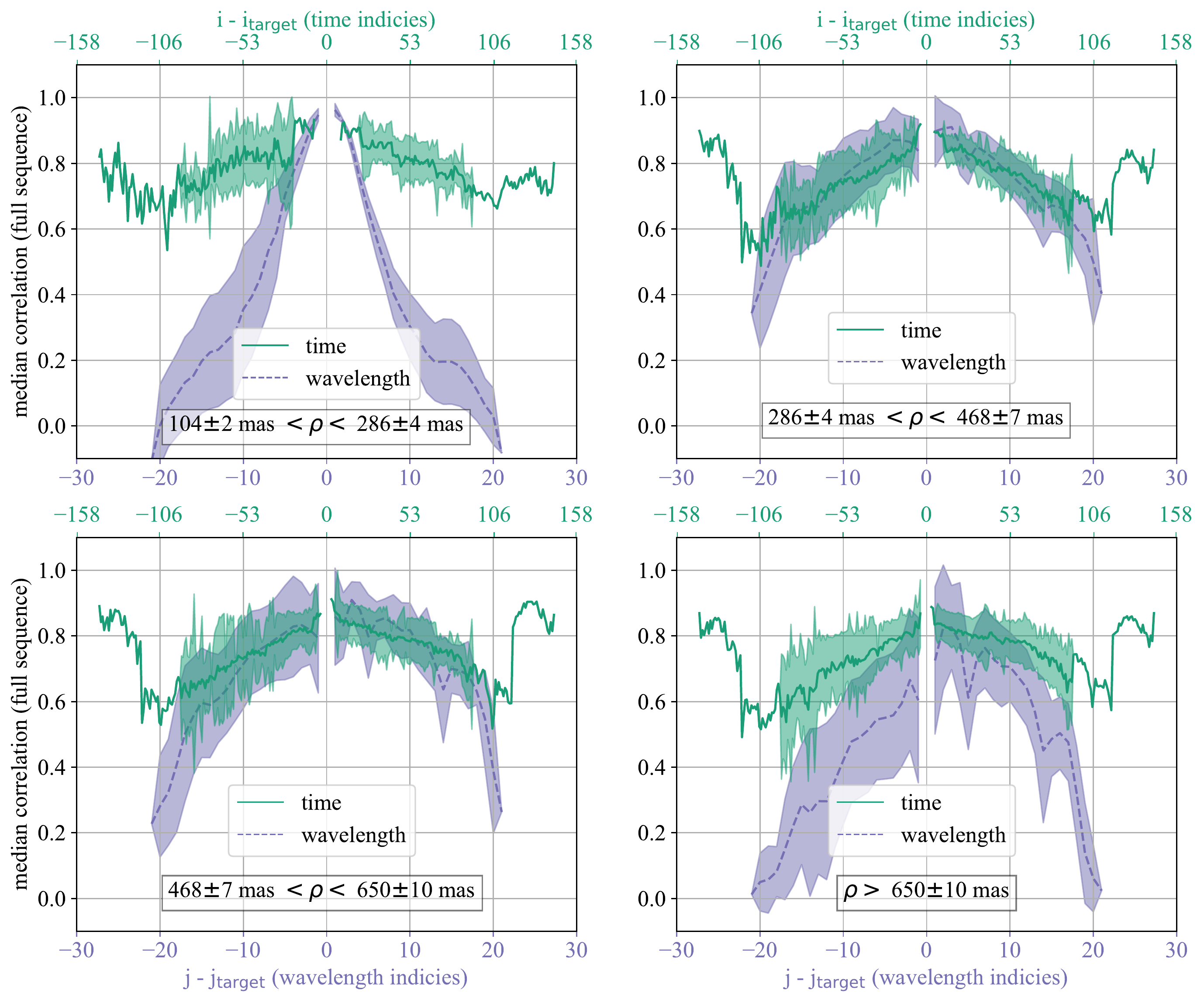}
\end{center}
\caption{The median correlation of selectable references over all the observed targets, separated into time and wavelength. The shaded regions show the robust standard deviation over all targets. Each panel corresponds to a different separation in the image. For every target image, each selectable reference image is identified as either the target wavelength and a different time ($j=j_\text{target},\; i\neq i_\text{target}$; top x-axis) or the target time and a different wavelength ($i=i_\text{target},\; j\neq j_\text{target}$; bottom x-axis). The main conclusion from this figure is that there is significant chromaticity at small and large separations over the full JHK bandpass; at adjacent wavelengths SDI may still be advantageous, but this figure suggests that this is not the case for a large wavelength lever arm across the full bandpass (see \S\ref{sec: results} for a validation of this hypothesis).}
\label{fig: cor_all}
\end{figure}

\begin{figure}[!ht]
\begin{center}
\includegraphics[width=1.0\textwidth]{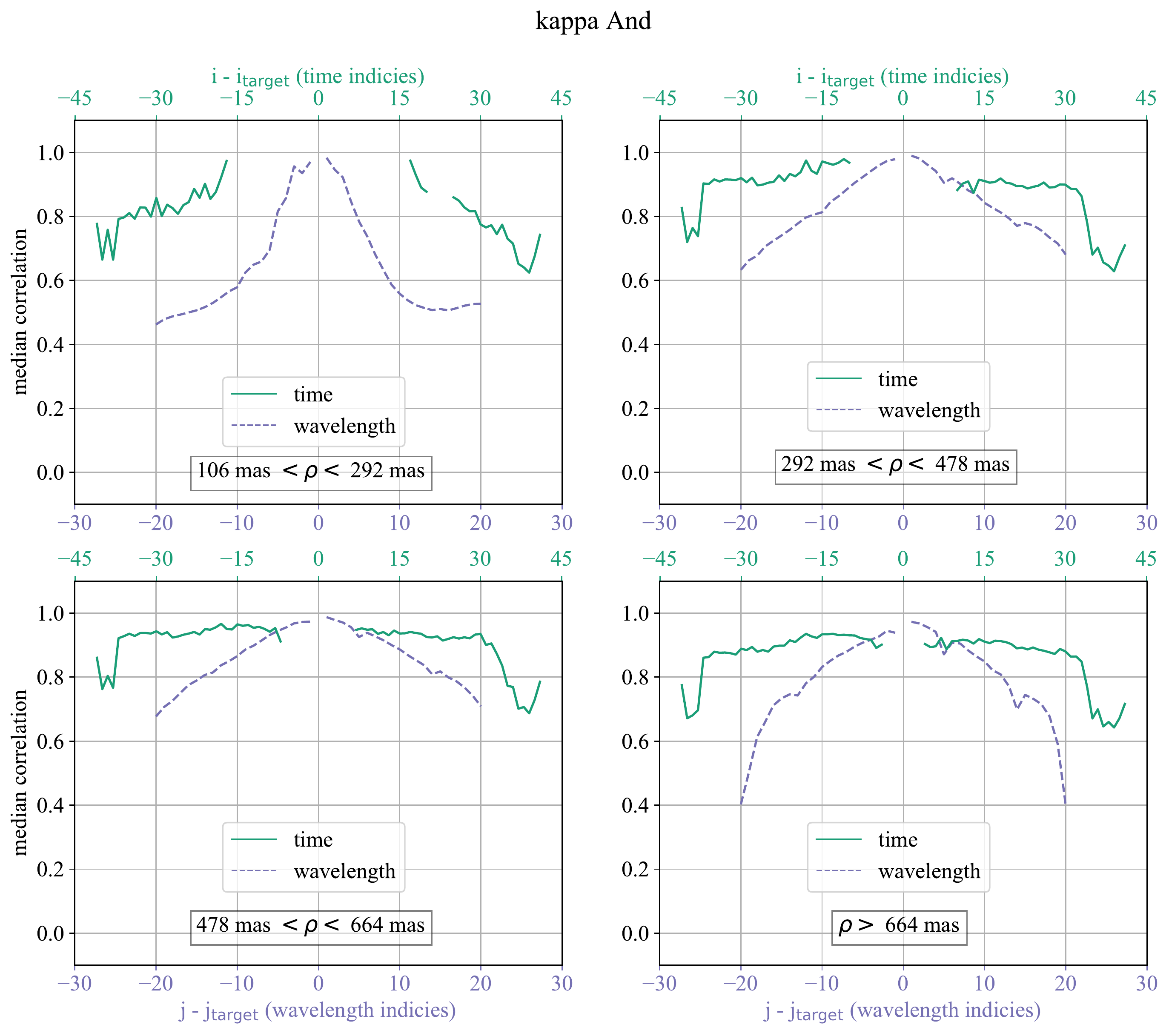}
\end{center}
\caption{Median correlation with time and wavelength for the $\kappa$ And dataset from \cite{kappa_and}, illustrating that on a more stable dataset chromaticity is generally also much better compared to Fig. \ref{fig: cor_all}.}
\label{fig: cor_seq}
\end{figure}

A generalization of Figure \ref{fig: cor1} extends to Figure \ref{fig: cor_all}, which shows the median correlation of all of our targets as a function of time and wavelength. 
Fig. \ref{fig: cor_all} is generated as follows:
\begin{enumerate}
\item For a single target image at a single time, wavelength, and user-defined separation, spectral type, and aggressiveness, a selection criterion is applied to all of the images in the oberving sequence; images that do not meet this criterion are discarded.
\item Correlation of the target image with every remaining image from step 1 is computed, normalized to unity using the standard deviation in both the target image and reference images. Steps 1-2 are illustrated in Fig. \ref{fig: cor1}.
\item In order to create a common zero-point for every subsequent target image (i.e., so the central index of Fig. \ref{fig: cor_all} is zero for every target image), each correlation value from step 2 is saved in a two dimensional array of size $2i_t\times2j_t$, where $i_t$ and $j_t$ are the total number of time stamps and wavelength slices, respectively (i.e., so both the negative and positive indices can be shown in Fig. \ref{fig: cor_all}). The $x$ and $y$ indices that fill a quarter of the total area of the two dimensional array for each correlation value from step 2 are $(2 i_t - i_\text{target})$ and $ (2 j_t - j_\text{target})$, respectively, where $i_\text{target}$ and $j_\text{target}$ are the indices of time and wavelength for the target image. This index definition ensures that the target image is in the center, or zero point, of each $2i_t\times2j_t$ array. Both the region removed by the selection criterion and the other three quarters of the array remain empty and are filled with ``not a number'' (NaN) placeholders.
\item Steps 1 - 3 are repeated over time and wavelength for every target image within a single observing sequence, generating a vector of two dimensional arrays of size $i_t j_t\times 2 i_t \times 2 j_t$. The autocorrelation of each target image is in the center of each $2 i_t \times 2 j_t$ array.
\item A median is taken across the zeroth axis of the vector in step 4 (i.e., the axis of dimension $i_t j_t$), ignoring NaN values, to generate a two dimensional array of size $2i_t\times2j_t$, representing the median correlation between the target image and every other image in the sequence. A slice along the $x$ direction of this array at $y=j_t$ (the zero point of the wavelength direction) reveals the median correlation of the target image (at $x=i_t$) vs. time, whereas a slice along the $y$ direction of this array at $x=i_t$ reveals the median correlation vs. wavelength.
\item Steps 1-5 are repeated over each user-defined spectral type. 
\item Steps 1-6 are repeated over each user-defined separation.
\item Steps 1-7 are repeated for each target, computing the median and standard deviation of correlation over time and wavelength for each separation.
\end{enumerate}
The outcome of the above procedure is illustrated in Figure \ref{fig: cor_all}, whereas steps 1-7 are illustrated in Fig. \ref{fig: cor_seq}, for which we used a separate dataset from our sample, the $\kappa$ And data from \cite{kappa_and}. Rather than a single target in our sample, the \cite{kappa_and} dataset was chosen for Fig. \ref{fig: cor_seq} to understand how the conclusions from Fig. \ref{fig: cor_all} change with higher quality AO correction. 

The main empirical conclusion from Figures \ref{fig: cor_all} and \ref{fig: cor_seq} is that SCExAO/CHARIS datasets are generally more correlated as a function of time than as a function of wavelength, most notably at small and large separations. This is particularly important at small separations, where PSF subtraction algorithms generally underperform compared to the rest of the image. For a given difference in parallactic angle and/or wavelength between a target and reference image, the respective azimuthal and/or radial movement of an exoplanet in between the two images is larger at larger separations. In other words, For a given aggressiveness, more reference images near the target image, both in time and wavelength, have to be discarded at lower separations compared to higher separations. Although a longer wavelength lever arm would help overcome this problem, Figures \ref{fig: cor_all} and \ref{fig: cor_seq} illustrate that this would generally add unnecessary noise in the covariance matrix inversion. A more correlated image at the same wavelength and different time will be selected instead. Later in \S\ref{sec: results} we will test the predictions from Fig. \ref{fig: cor_all} on our PSF subtraction pipeline. 

With that said, at small separations in Fig. \ref{fig: cor_all} and at all separations in Fig. \ref{fig: cor_seq}, reference images are actually more correlated at the closest wavelengths than at the closest times. Although these images will be preferentially selected for SDI processing, this approach will not necessarily reach the deepest (throughput-corrected) final contrasts. Selecting a closer wavelength will reach a better non-throughput-corrected contrast but will also cause more self-subtraction, therefore requiring a larger throughput correction. Thus, the optimal reference selection based on final, throughput-corrected contrast will ultimately depend on optimizing the trade off between correlation and self-subtraction, as in \cite{me_gpi}. Additionally, although the chromaticity is still worse than the temporal-stability in Fig. \ref{fig: cor_seq}, note that there is a significant improvement in both temporal stability and chromaticity between Fig. \ref{fig: cor_all} and \ref{fig: cor_seq}. Although the improvement from temporal stability is most likely from improved AO performance (strehl ratios for the \citealt{kappa_and} dataset are around 0.92), the origin of chromaticity improvement is not clear and will require further investigation and analysis, beyond the discussion below.

Although the specific origin of the observed chromaticity is beyond the scope of this paper, we discuss a few possibilities here that could explain these results. In general, Fresnel propagation generates a chromatic evolution of the wavefront from any out-of-pupil plane optics, particularly for optics that are transmissive and/or near the focal plane \citep[e.g.,][]{gpi_fresnel}, as well as from atmospheric scintillation \citep{guyon_limits}, although both of these effects are expected to be very weak at the current level of raw contrasts (but see \citealt{alex}, who show that in some cases scintillation effects are observed in GPI data). Additional sources of the observed behavior may be algorithmic in nature, arising from either numerical interpolation errors and/or DRP extraction errors. These numerical interpolation errors are wavelength-dependent when aligning data cubes, since, as discussed in \S\ref{sec: drp}, the full CHARIS bandpass goes from near the Nyquist sampling limit in J band to more oversampled in K band. Although numerical interpolation errors are more likely to disproportionally effect the J band (from under-sampling effects) while DRP extraction errors are more likely to affect the K band (from increased thermal background effects), further investigation is needed to understand the relative impact of either effect. At small separations there are two additional possible origins of the observed chromaticity: the atmospheric dispersion corrector (ADC) and the FPM IWA. The ADC does not sufficiently offset differential atmospheric tip/tilt to center the FPM on the star at the red edge of the K band. For the FPM IWA, because our chromaticity analysis uses magnified data cubes to align speckles as a function of wavelength, the IWA on magnified data cubes varies between about 2 and 4 $\lambda/D$ at the red and blue ends of the bandpass, respectively, which could explain the observed chromaticity in the inner-most radial zone.
\section{Contrast Curves}
\label{sec: results}

\begin{figure}[!h]
\centering
\includegraphics[width=1.0\textwidth]{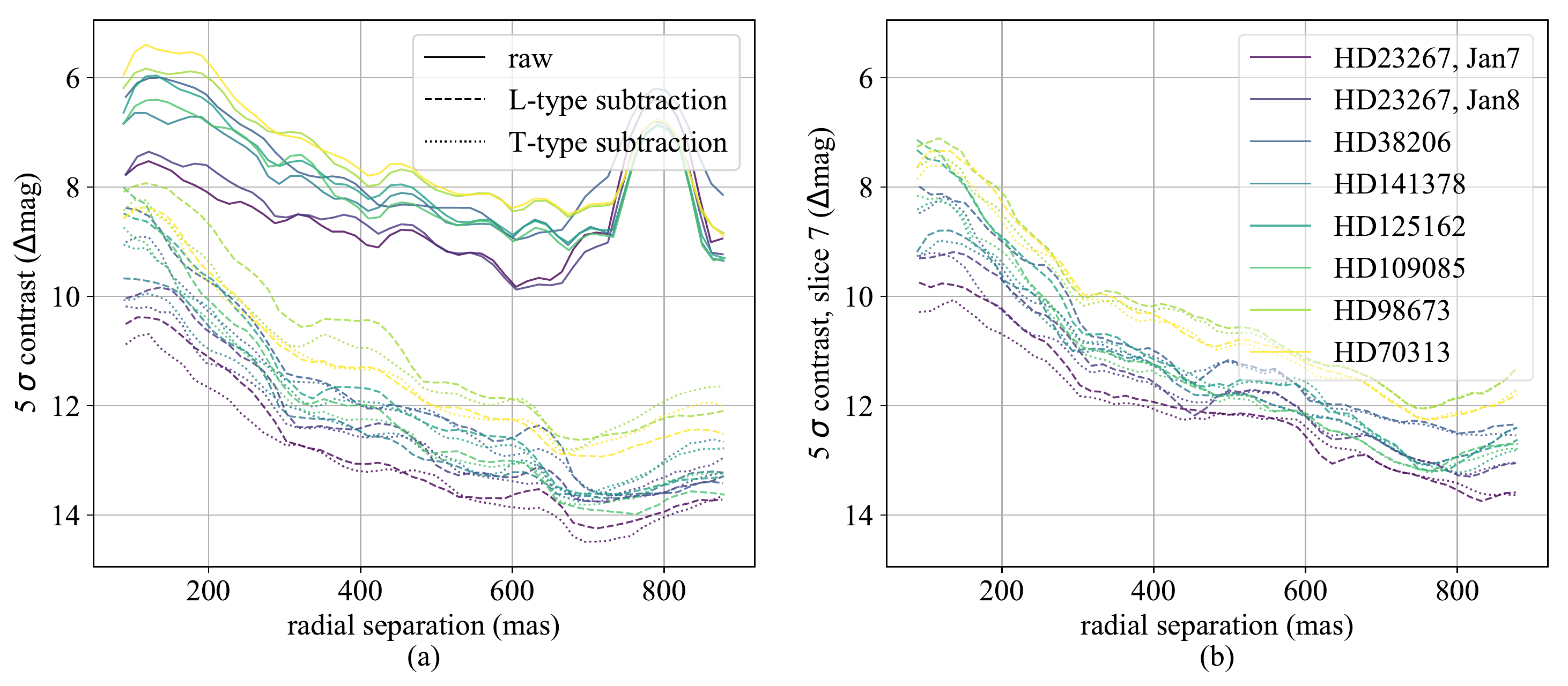}
\caption{(a): contrast curves for all of our targets. (b): Final contrasts at a single wavelength, slice 7 = 1.5 $\mu$m. Each target is color coded (the same in both figures); the solid, dashed, and dotted lines indicate the raw, final L-type, and final T-type contrasts, respectively.}
\label{fig: contrast_curves}
\end{figure}

Our 5$\sigma$ contrast curves are shown in Fig. \ref{fig: contrast_curves}. Final contrasts are calculated using the standard deviation at each separation in a 3 pixel-wide annulus on the collapsed, back-rotated images. Raw contrasts are calculated on a median image of the stack of the registered, magnified, flux-normalized, high-pass-filtered images (i.e., the same image preparation used before the correlation analysis in addition to a median stack of the full sequence). Also note that the frames used to calculate raw contrast are not de-rotated to align the parallactic angle, thus keeping quasi-static aberration aligned during the median collapse.

Although other deeper contrasts have been reached with CHARIS on more stable datasets \citep[e.g.,][]{kappa_and, evan}, the main purpose of this paper is instead to provide a detailed correlation analysis (\S\ref{sec: corr}) that can be used as a reference for future CHARIS observations (e.g., considering the impact of SDI in broadband mode). 

With that said, Fig. \ref{fig: contrast_curves} a and b does show that T-type subtractions generally reach up to $\sim$50\% deeper contrasts than the corresponding L-type subtractions for the same target, illustrating that SDI \textit{is} improving contrast. Contrasts in a single slice are shown in addition to the final broadband contrasts to control for the effects of noise averaging in comparing an L- vs. T-type reduction. In applying a weighted average using the assumed planet spectral type to collapse the final subtracted cube, noisier frames that lie further away from the bright part(s) of the spectrum are given less weight to the final collapsed cube \citep{tloci}, and so looking only at a single wavelength slice at the peak of the H band T-type spectrum removes this affect. Although other wavelengths will not show as much of a performance gain further away from the T-type spectrum peak, a deeper contrast reached by a T-type vs. L-type, at any wavelength, illustrates that SDI is being used to improve contrast. This improvement occurs because, within the 15 reference images that are chosen to subtract a target image at a given time and wavelength, some are at the same time but different wavelength as the target image. These results are consistent with our analysis in \S\ref{sec: corr} which showed that chromaticity generally prevents using a reference image for SDI with a large wavelength lever arm, but that reference images closer in wavelength were often more correlated than images closer in time. For a T-type subtraction, the relative lack of long dark shadows in the spectral (i.e., radial) direction of the right panel of Fig. \ref{fig: psf_subt} is also consistent with these results; selecting the reference images closest in wavelength, especially for a T-type subtraction, are still often more correlated than the closest images in time. 

Along these lines, because the main conclusions of Figures \ref{fig: cor_all} and \ref{fig: cor_seq} indicate that reference images at a large wavelength difference from the target image are unlikely to be selected (i.e., separate from the hypothesis of whether or not SDI is improving contrast), we carried out the exact same PSF subtraction procedure as described in \S\ref{sec: psf_subt} on all seven targets using only the H band wavelengths within the broadband data cubes (i.e., slices 6 - 13). The idea here is to see what level of contrast improvement is lost by removing the broadband feature of CHARIS while still using SDI in a standard $\sim$20\% bandpass. Note that using the same parameters as in \S\ref{sec: psf_subt} (e.g., N$_\text{ref}$=15) means that there are less available reference images as a function of wavelength to select for the H band-only sequence. However, because reference images are selected by correlation, our chromaticity analysis in \S\ref{sec: corr} and Figure \ref{fig: cor_all} has determined that images at these wider wavelength differences are generally not selected, and so we expect this discrepancy to have a negligible effect on final contrast. 
\begin{figure}[!h]
\centering
\includegraphics[width=0.65\textwidth]{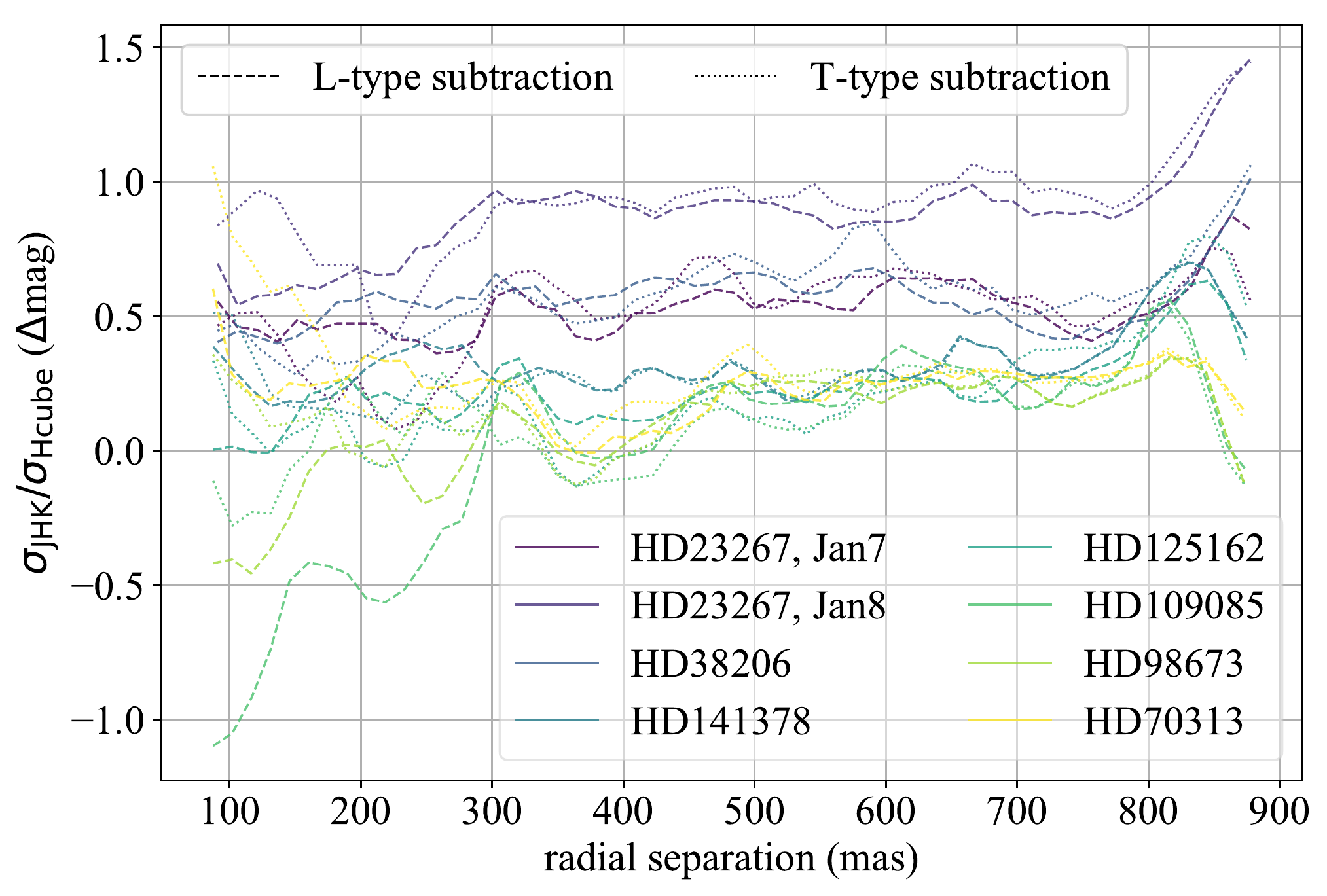}
\caption{The difference in contrast for each target when using the full JHK dataset vs. only the H band to do PSF subtraction, illustrated here for a single wavelength slice at the center of H band (slice 11 in the full JHK cube). The color and line styles are the same representation as in Fig. \ref{fig: contrast_curves}.}
\label{fig: h_only}
\end{figure}

Fig. \ref{fig: h_only} illustrates the contrast difference on a single slice in the center of H band for all of our targets between the full broadband mode and the H band mode. Note that comparing the collapsed cubes between the broadband and H-only subtractions would introduce a systematic bias; in the final wavelength collapse the broadband dataset would average more slices than the H-only dataset, producing a deeper final contrast for the former even if the individual subtracted cubes are the same for both sequences. In general, all of our April 2017 observations show little to no gain using the broadband dataset, as expected from Fig. \ref{fig: cor_all}. Interestingly, the January 2018 observations, which reach the deepest raw and final contrasts in Fig. \ref{fig: contrast_curves}, instead consistently show a $\sim$0.5 - 1 magnitude gain from using the broadband mode. Also when using the broadband mode, in a few cases at separations less than about 200 mas, the L-type subtractions appear to decrease in contrast by up to a magnitude.

Ultimately, the origin of this observed chromaticity that may be improving and/or degrading contrast will need to be better understood before a detailed analysis leading to mitigation strategies can be done. As discussed in \S\ref{sec: corr}, many different possible sources of chromaticity could reproduce the results in Figures \ref{fig: cor_all}, \ref{fig: cor_seq}, and \ref{fig: h_only}. These sources may be algorithmic and/or optical in nature, and could be quasi-static and/or time-variable. However, independent of the origin, in this paper we illustrate that chromaticity in our broadband datasets is limiting the final achievable contrast, thus motivating the need for further work towards more achromatic ExAO systems. 

\section{Conclusions}
\label{sec: conclusions}
In this paper, we have presented a detailed correlation analysis, data processing architecture, and final contrast curves of our SCExAO/CHARIS broadband datasets to the community. Our main findings are as follows:
\begin{enumerate}
\item\label{pt: chromaticity} A correlation analysis of the SCExAO/CHARIS system across all seven of our targets shows that for small ($\lesssim$300 mas) and large ($\gtrsim$650 mas) separations, chromaticity across the full JHK bandpass is worse than temporal stability across the full observing sequence (Fig. \ref{fig: cor_all}). This result suggests that at these separations, the large wavelength ``lever arm'' of our CHARIS broadband datasets could not be used advantageously to improve SDI post-processing (i.e., slices in J band can generally not be used to subtract slices in K band, and vice versa).
\item\label{pt: t-type} With that said, we find that our T-type PSF subtractions generally reach up to $\sim$50\% deeper final contrasts than our L-type subtraction of the same target (Fig. \ref{fig: contrast_curves}), illustrating that more correlated wavelength slices close to the target image wavelength are being used to improve contrast. This is consistent with the findings of point \ref{pt: chromaticity} above (i.e., Fig. \ref{fig: cor_all}); together, these results illustrate that improved contrast from SDI processing of our targets comes from selecting reference slices closer in wavelength to the target image where chromaticity is less significant.
\item We present a modified framework for exoplanet algorithmic throughput correction using forward modelling that is designed for the CHARIS broadband mode (\S\ref{sec: psf_subt}). As a result, simulated planets are recovered in good agreement with their input values even when close to the 5$\sigma$ noise floor (Fig. \ref{fig: boot}).
\item PSF-subtracted sequences using only the H-band slices within the broadband dataset generally reach final contrasts that are similar to the final contrasts of the same slices in the broadband sequences (Fig. \ref{fig: h_only}), validating the claim made in points \ref{pt: chromaticity} and \ref{pt: t-type} above.
\end{enumerate}

Beyond the work presented in this paper, as long as ExAO systems remain limited by temporal stability (i.e., preventing subtraction to the photon noise limit using ADI and/or RDI post-processing, thus motivating the additional need for SDI-processing), chromaticity will also limit the deepest contrasts that can be reached. Looking towards the future, coherent differential imaging \citep[CDI; e.g.,][and references therein]{guyon_cdi, speckle_nulling, scc_orig, efc, wallace, phase_diversty, cdi} is a complimentary approach towards reaching deeper contrasts on ExAO systems. CDI can operate in a narrow bandpass (i.e., in contrast with SDI) and is therefore, in principle, not limited by chromaticity. However, detailed characterization of colder/lower mass exoplanet atmospheres will continually push the ExAO community towards reaching deeper broadband contrasts, in which case chromaticity will ultimately become a fundamental limitation to this goal. The future push towards reaching deeper contrasts will therefore also require a push towards implementing more achromatic broadband systems.
\section*{Acknowledgements}
The data in this paper was obtained through 2017B open use Canadian and US Gemini exchange time \citep{scexao_proposal}. We gratefully acknowledge research support of the Natural Sciences and Engineering Council of Canada through the Postgraduate Scholarships-Doctoral discovery grant (DG), and Technologies for Exo-Planetary Science Collaborative Research and Training Experience programs. The development of SCExAO was supported by the JSPS (Grant-in-Aid for Research \#23340051, \#26220704 \#23103002), the Astrobiology Center of the National Institutes of Natural Sciences, Japan, the Mt Cuba Foundation and the directors contingency fund at Subaru Telescope. CHARIS was built at Princeton University under a Grant-in-Aid for Scientific Research on Innovative Areas from MEXT of the Japanese government (\# 23103002). We thank Jason Wang and Jean-Baptiste Ruffio for helpful discussions and suggestions about forward modelling. The authors thank the anonymous referee for his or her comments and suggestions that have significantly improved this manuscript. The authors acknowledge and support the cultural and spiritual importance of the summit of Mauna Kea to the Hawaiian native community. We are grateful to share and respect this land with the community.
\appendix
\section{Additional Correlation Analysis}
\label{sec: cor_cut}
\begin{figure}[!h]
\centering
	\begin{minipage}[b]{0.4\textwidth}
		\begin{center}
		\includegraphics[width=1.0\textwidth]{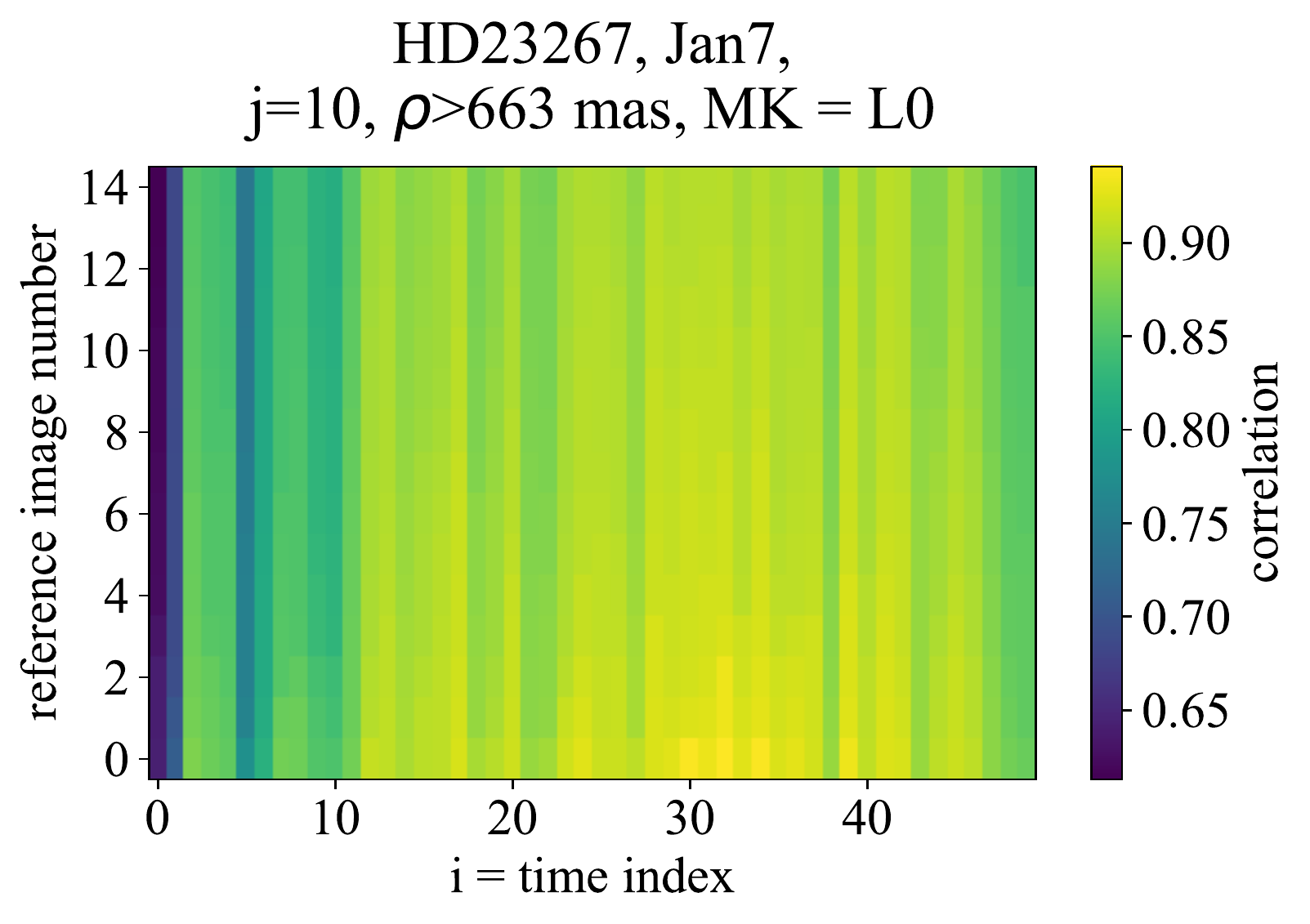}
		(a)
		\label{fig: b}
		\end{center}
	\end{minipage}	
	\begin{minipage}[b]{0.4\textwidth}
		\begin{center}
		\includegraphics[width=1.0\textwidth]{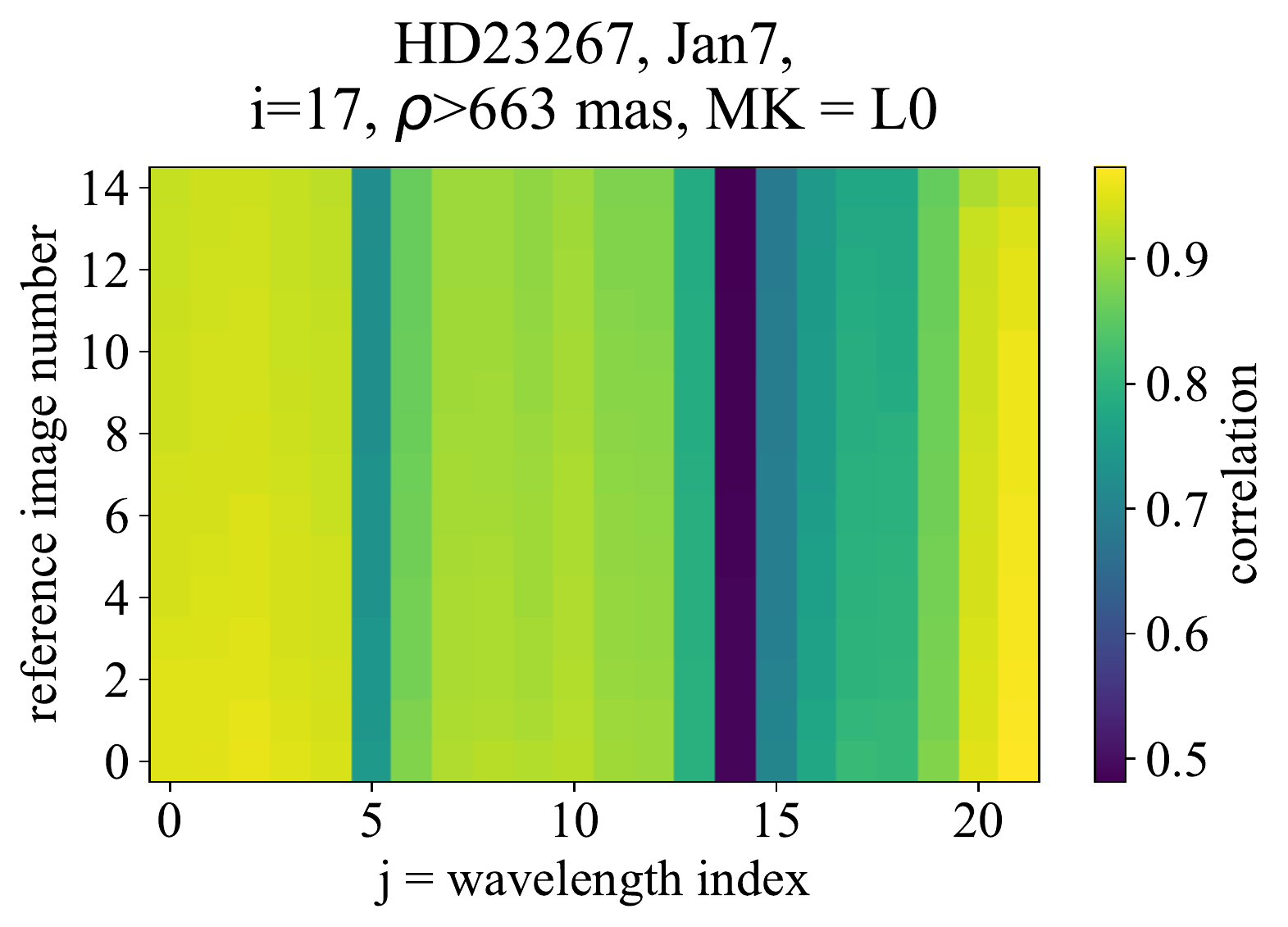}
		(b)
		\label{fig: c}
		\end{center}
	\end{minipage}	
	\begin{minipage}[b]{0.4\textwidth}
		\begin{center}
		\includegraphics[width=1.0\textwidth]{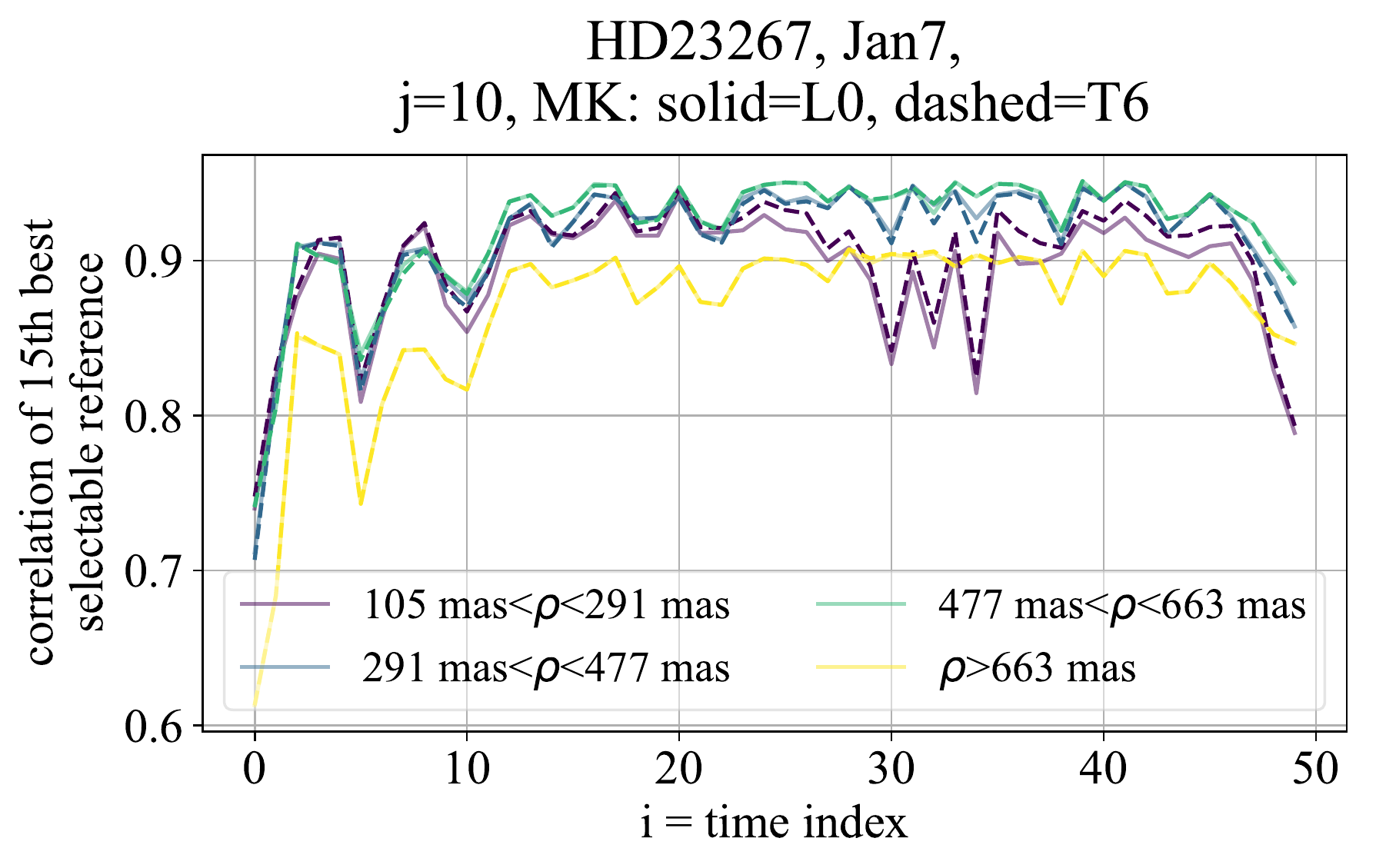}
		(c)
		\end{center}
	\end{minipage}
	\begin{minipage}[b]{0.4\textwidth}
		\begin{center}
		\includegraphics[width=1.0\textwidth]{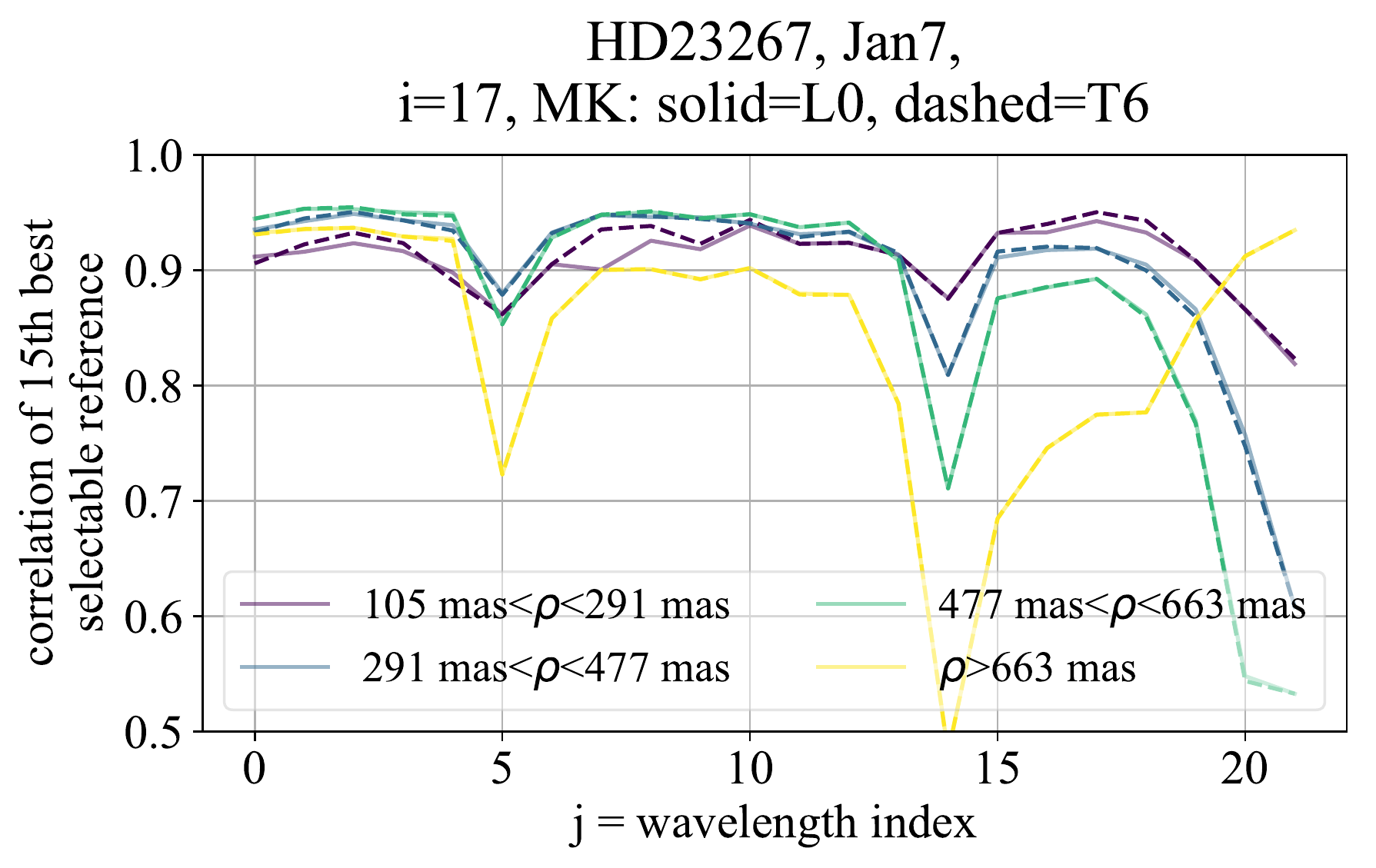}
		(d)
		\label{fig: d}
		\end{center}
	\end{minipage}	
\caption{(a) and (b): the 15 most correlated selectable reference images (i.e., reference image numbers 0 through 14,  using a zero-index counting system) at a single wavelength slice as a function of time and at a single time stamp as a function of wavelength, respectively, both for large separations and assuming a L0 template selection criterion. (c) and (d): The 15th most correlated selectable reference image at all separations and for two input spectral types (c) at a single wavelength as a function of time and (d) at a single time as a function of wavelength (i.e., reference image number 14 as a function of time and wavelength).}
\label{fig: cor2}
\end{figure}
\begin{figure}[!h]
\centering
\includegraphics[width=0.59\textwidth]{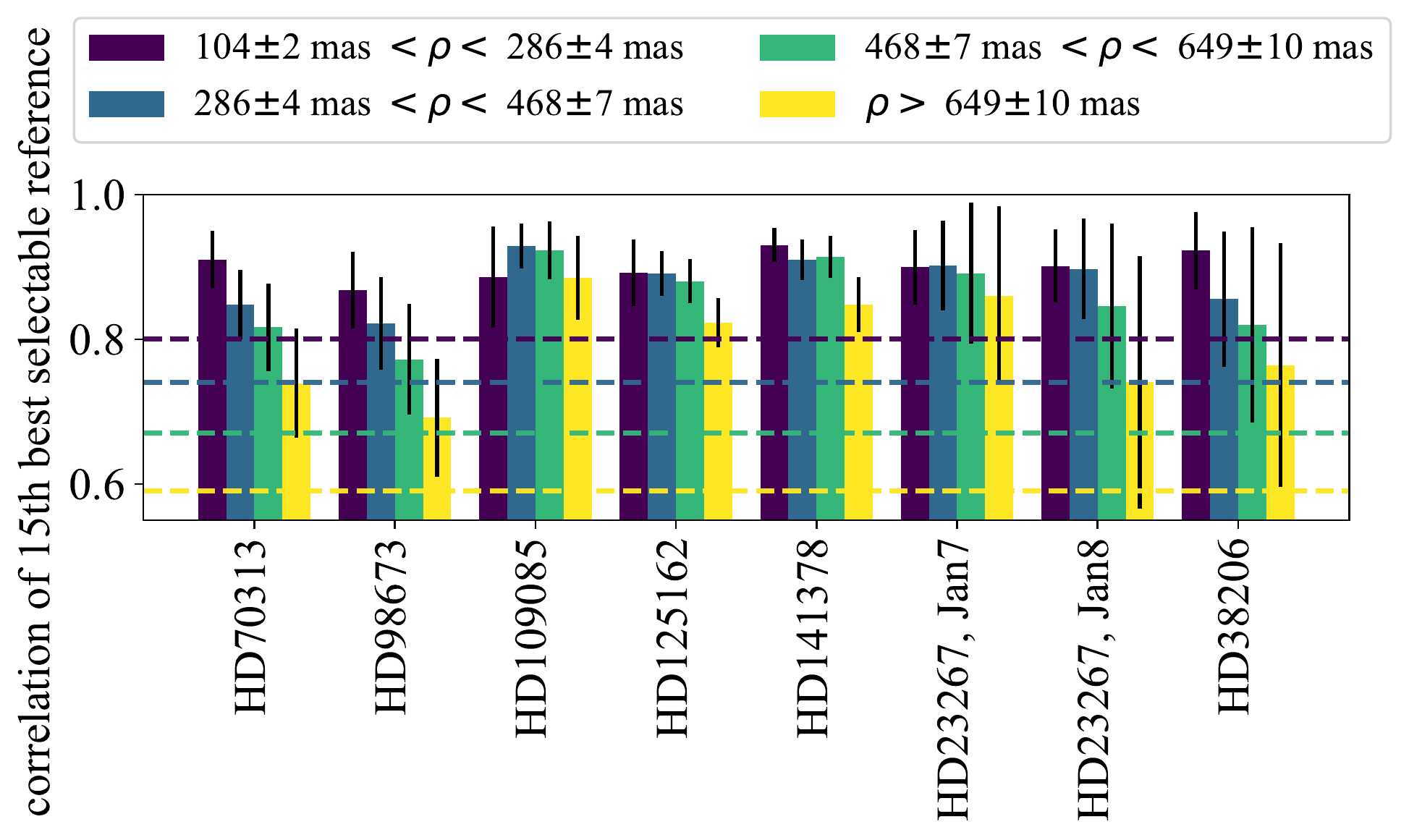}
\caption{The median correlation of the 15th most correlated, selectable reference image and corresponding robust standard deviation over all times, wavelength, and input spectral types, at each separation for each target. Our adopted correlation cut values are illustrated by the horizontal dashed lines, color coded to each separation.}
\label{fig: cor3}
\end{figure}
Additional correlation analysis informed our choice of values for a correlation cut, as discussed in \S\ref{sec: psf_subt}. Still using the HD 23267 Jan. 7 dataset, frame 17, slice 10 as an example (as in Fig. \ref{fig: cor1}), Fig. \ref{fig: cor2} illustrates the trends in correlation of selected reference images at either the same time or same wavelength (i.e., i = 17 or j = 10, respectively). The top two panels show the correlation trends of the 15 most correlated images as defined in \S\ref{sec: psf_subt}, N$_\text{ref}$=15) at large separations as a function of time (a) or wavelength (b) at large separations, using an aggressiveness of 0.5, and using an assumed L0 input spectrum. Uncorrelated frames can be identified as vertical ``streaks" in these images (e.g., the water bands at j = 5 and j = 14, as in Fig. \ref{fig: cor1}). The chosen correlation cut value at this separation should reject these uncorrelated frames from the sequence in order to prevent unnecessary noise amplification in the least-squares algorithm \citep{sosie}. Because we require that all 15 reference images lie above our defined correlation cut value, the bottom panels of Fig. \ref{fig: cor2} show a one dimensional slice of the top two panels: the correlation of reference image number 14 (i.e., the 15th most correlated selectable reference, using a zero-index counting system) as a function of time (c) and wavelength (d) for both spectral types and all separations. Similarly, at all separations and spectral types we would like to discard the frames where there are ``dips'' in these plots in order to optimally reject uncorrelated images and avoid propagating them through the least-squares algorithm. Fig. \ref{fig: cor2} c and d also suggest that larger separations may be less correlated overall and that a different correlation cut value will be needed for each separation. Note that the low correlation at larger separations may be a result of wavefront decorrelation at those spatial frequencies and/or increased background noise because the speckles at those separations are detected at a lower S/N. At a given separation, using a correlation cut value that is too high will reject too many images at that separation and return only a minimal number of subtracted frames to the final collapsed cube; the ``optimal'' correlation cut value should balance rejecting too many images with propagating too much noise through the least-squares algorithm.

The extension of Fig. \ref{fig: cor2} c and d is to calculate the median behavior over a full sequence and for all targets. Thus, for a single target at a given separation, in Fig \ref{fig: cor3} we compute the median and robust standard deviation for the 15th most correlated reference image over all target images (time and wavelength) and input spectral types. Fig. \ref{fig: cor3} illustrates that, for most targets, the 15th most correlated image decreases in correlation with separation, motivating the use of different correlation cut values at each separation. The horizontal dashed lines in Fig. \ref{fig: cor3} illustrate our chosen correlation cut values at each separtion. We set these values to generally lie near the bottom of the corresponding 1$\sigma$ error bars over all targets. Using this definition to define the correlation cut values will balance the rejection of bad frames relative to the median correlation value without discarding too many frames from the sequence. In general, the optimal correlation cut value will vary for each least-squares subtraction and is degenerate with SVD cutoff and number of reference images \citep{sosie}; this will ultimately require a more optimized approach \citep[e.g.][]{me_gpi}, but is beyond the scope of this paper.

\bibliography{refs}

\end{document}